# Phase transition at finite temperature in one dimension: Adsorbate ordering in Ba/Si(111)3×2

Steven C. Erwin and C. Stephen Hellberg

*Center for Computational Materials Science, Naval Research Laboratory,*

*Washington, D.C. 20375*

## Abstract

We demonstrate that the Ba-induced Si(111)3×2 reconstruction is a physical realization of a one-dimensional antiferromagnetic Ising model with long-range Coulomb interactions. Monte Carlo simulations performed on a corresponding Coulomb-gas model, which we construct based on density-functional calculations, reveal an adsorbate-ordering phase transition at finite temperature. We show numerically that this unusual one-dimensional phase transition should be detectable by low-energy electron diffraction.

Keywords:  Density functional calculations;  Monte Carlo simulations;  Equilibrium thermodynamics and statistical mechanics;  Ising models; Low energy electron diffraction;  Surface thermodynamics; Silicon; Alkaline earth metals

Corresponding author:
Dr. Steven C. Erwin
Center for Computational Materials Science
Naval Research Laboratory
Washington, DC, 20375-5000
Tel 202-404-8630
Fax 202-404-7546
Email erwin@dave.nrl.navy.mil

Can a one-dimensional system exhibit a phase transition at finite temperature? The answer depends on the nature of the interactions, in particular their range and sign. For Heisenberg systems, Mermin and Wagner proved that when the interactions are of either sign and short-ranged, for example monotonically decaying with distance faster than $R^{-3}$, there can be no such phase transition [1]. This result has been extended to include interactions monotonically decaying faster than $R^{-2}$ and certain oscillatory interactions decaying faster than $R^{-1}$ [Ref. 2]. For both Heisenberg and Ising systems with ferromagnetic interactions of the form $J(R)=R^{-\alpha}$, a phase transition exists only in the range $1 < \alpha \leq 2$ [Ref. 3]. For purely antiferromagnetic interactions, no analytical results have yet been obtained.

Here we identify and study a physical realization of the one-dimensional antiferromagnetic Ising model with long-ranged $R^{-1}$ interactions: the Ba-induced Si(111)3×2 reconstruction. We first use density-functional theory (DFT) to show that the subsystem of Ba adsorbates can be accurately represented as a Coulomb gas of ions confined to one-dimensional channels. We compute the statistical properties of this Coulomb gas using a Monte Carlo method, and identify an adsorbate-ordering phase transition near room temperature. We demonstrate that this phase transition should be detectable by low-energy electron diffraction (LEED). Finally, we show that this Coulomb-gas model is equivalent to an antiferromagnetic Ising model with $R^{-1}$ interactions. Thus, Ba/Si(111)3×2 is a realization of a model whose properties narrowly escape being subject to the above theorems.

Ba/Si(111)3×2 is a period-doubled variant of the 3×1 "honeycomb-chain channel" (HCC) reconstruction induced by alkali adsorbates [4,5,6]. In the 3×1 HCC, the Si atoms in the surface layer form one-dimensional honeycomb chains, separated by channels that accommodate the adsorbates. Within a channel, there are two nearly degenerate adsorption sites: $T_4$ (above a third-layer Si atom) and $H_3$ (above a fifth-layer Si atom). The adsorbates donate their electrons to surface Si orbitals. When the alkali coverage is 1/3 monolayer, there are exactly enough electrons available to completely fill the bonding combinations of Si orbitals, leaving the antibonding combinations empty. Fulfilling this electron counting principle confers stability to the 3×1 HCC.

For divalent adsorbates such as Ba, the same electron counting principle requires the coverage to be halved, to 1/6 monolayer [7]. At zero temperature, the ground state configuration of adsorbates has doubled periodicity along the chain direction. For Ba, it is most favorable to occupy every second $T_4$ site, as shown in Fig. 1. Placing the adsorbates in every second $H_3$ site raises the total energy only very slightly (by 35 K per Ba within our DFT calculations, described below). Lee *et al.* [8] have obtained similar results, finding that the $T_4$ site is preferred and the $H_3$ site is slightly (~100 K) higher in energy. Electronic structure calculations confirm that, for either configuration, both valence electrons of Ba are donated to surface Si orbitals.

To investigate the statistical properties of Ba/Si(111) at finite temperature, we first construct a model representing the Ba adsorbate subsystem. The interactions within this model have a simple analytical form whose coefficients we obtain from DFT calculations on the full Ba/Si(111) system. The DFT calculations were performed in a slab geometry consisting of eight layers of Si plus the reconstructed surface layer; all atomic positions were relaxed except for the bottom double layer, which was passivated with hydrogen. Total energies and forces were calculated within the generalized-gradient approximation, using projector-augmented-wave potentials for Ba and ultrasoft pseudopotentials for Si [11,12]. The plane-wave cutoff was 150 eV (with convergence checks using 250 eV), and the 3x2 unit cells used a 4x8 sampling of the surface Brillouin zone; these values were sufficient to converge all energy differences to less than 1 meV.

The total energy of a configuration of adsorbates at locations $\mathbf{R}$=(x,y) (the variations in z are small compared to those in x and y, so we approximate the system by its 2D projection onto the surface plane) is given by a local interaction with the substrate, plus a screened Coulomb interaction between the ionized adsorbates of charge Q,

$$E = \sum_i V_{sub}(x_i) + \sum_{ij} Q^2/\varepsilon|\mathbf{R}_i - \mathbf{R}_j|, \qquad (1)$$

where $x_i$ is the projection of the $i^{th}$ adsorbate's position $\mathbf{R}_i$ along the chain direction, Q=+2 is the nominal ionic charge of the Ba adsorbates, and $\varepsilon$ is an effective dielectric constant which we determine below. We evaluate the substrate interaction by computing the DFT total energy for Ba positions between the $T_4$ and $H_3$ sites, relaxing all other atomic coordinates within 3×2 periodicity. The variation of the total energy along the projected coordinate is shown as the circles in Fig. 2. A single barrier separates the metastable $T_4$ and $H_3$ sites, and the DFT energies can be accurately fit to the form $V_{sub}(x)$=A[1 - cos(2πx/a)] + B[1 - cos(4πx/a)], where a is the surface lattice constant; our fit gives A=6 K and B=794 K. (For comparison, Lee *et al.* [8] find A=58 K and B=841 K.)

Since our model assumes a purely Coulombic interaction between adsorbates, the second term in Eq. (1) is entirely specified once $\varepsilon$ is determined. We do this by computing the DFT total energy (the squares in Fig. 2) as the Ba-Ba spacing is varied, while relaxing all other degrees of freedom. To do this, we used a doubled unit cell containing two Ba ions. We constrained one Ba ion at its equilibrium $T_4$ site, and varied the position of the other between its equilibrium $T_4$ site and the neighboring $H_3$ site, while relaxing all other atoms in the cell. For small changes in the spacing the energy is dominated by the substrate interaction, but as one adsorbate approaches the $H_3$ site the contribution from Coulomb interactions becomes comparable to the substrate barrier. Thus, we can determine $\varepsilon$ numerically by a least-squares fit of Eq. (1) to the energies in Fig. 2, with the result $\varepsilon=7.9$. It is clear from the quality of the fit that the model interactions accurately reproduce the DFT results for the full system.

The geometry of our model is a two-dimensional planar array of straight lines representing the adsorbate channels. The separation between the channels of Ba/Si(111) is sufficiently small that the average interparticle separation – and therefore the Coulomb interaction – between Ba ions within a channel is comparable to that across channels. Nevertheless, we will show below that Ba/Si(111) behaves very much like a one-dimensional system. More precisely, we will show that the thermodynamic behavior of our two-dimensional model of Ba/Si(111) is very similar to that of its one-dimensional analog: Ba ions moving within a single, isolated channel. Qualitatively, this is not surprising, since the <u>change</u> in the energy of an adsorbate as it moves from a $T_3$ to a $H_4$ site is is dominated by interactions with other adsorbates along the direction of motion – i.e. within the same channel. Quantitatively, we find that the specific heat, structure factor, and finite-size scaling of the estimated transition temperature are nearly identical for the 1D and 2D models. Moreover, the predicted thermodynamic critical temperatures in the 1D and 2D models are essentially identical.

We use the Metropolis algorithm to find the equilibrium distribution of adsorbates as a function of temperature. We approximate the infinite surface by periodic supercells containing N adsorbates, and then extrapolate the results to the thermodynamic limit, N → ∞, using a method described below. In the 2D model, the supercell lattice vectors have the form $\vec{L}_1 = 2a(N/4)\,\hat{x} - b\,\hat{y}$, $\vec{L}_2 = 4b\,\hat{y}$, where $\hat{x}$ and $\hat{y}$ are unit vectors along and normal to the adsorbate lines, respectively, and b = $(3\sqrt{3}/2)$a is the spacing between lines. This choice leads to the largest possible value, N, for the periodicity along a single line, and hence maximizes the allowed range of thermal fluctuations in the local particle density within a line. In the 1D model, the single lattice vector is simply $\vec{L} = 2aN\,\hat{x}$.

Traditional Metropolis algorithms become extremely inefficient near phase transitions due to high free-energy barriers that must be overcome. To speed convergence, we use the parallel tempering method [13], a Monte Carlo technique in which many configurations, spanning a wide range of temperatures, are separately simulated using traditional Metropolis sampling. Configurations at adjacent temperatures are occasionally swapped, maintaining detailed balance. Here, we simulate 52 temperatures simultaneously, between 100 K and 4000 K. We find that the temperature of each configuration changes many times during the simulation, indicating efficient sampling of phase space.

In Fig. 3a we show the computed specific heat, $C_V = (\partial E/\partial T)_V$, as a function of temperature, in the 1D and 2D models for N=25. The behavior is very similar for both, and suggests a phase transition near 300 K, with the expected singularity in $C_V(T)$ softened by finite-size effects. In Fig. 3b we plot the structure factor, S(q) = $\langle \rho(q)\rho(-q)\rangle/N$, where $\rho(q) = \sum_j \exp(iqx_j)$, for the ordering wavevector q=π/a. The 2D and 1D results for N=25 are again very similar. In the 1D model, large values of N lead to a sharp transition around 300 K.

To determine the critical temperature in the thermodynamic limit, we compute the fourth-order cumulant $U_L(T) = <(\rho(q)\rho(-q))^2>/<\rho(q)\rho(-q)>^2 - 1$. The transition temperature is given by the intersection of $U_L(T)$ for different system sizes [9,10]. The resulting transition temperatures for all possible pairs of system sizes $(N_1, N_2)$ are plotted in Fig. 3c as a function of $<N> = (N_1 N_2)^{1/2}$. We find that both the 2D and 1D results exhibit the simple finite-size scaling shown, $T_c = T_0 + T_1/<N>$. We identify $T_0$ as the thermodynamic limit of the critical temperature, and thereby find $T_c = 270$ K for both the 2D and 1D models.

Above the critical temperature, we expect LEED data to reflect the loss of 3×2 order of the Ba subsystem while still showing the 3×1 order of the underlying Si substrate. We demonstrate this effect by calculating LEED spot profiles using the equilibrium Ba positions from our Monte Carlo simulations. We used a parallelized version of the Barbieri/Van Hove SATLEED package [14], which accounts for multiple scattering from the Ba adsorbates and from the reconstructed Si substrate, which we take to have ideal 3×1 periodicity. In our finite-size Monte Carlo simulations the loss of long-range order occurs gradually, as is evident from the temperature dependence of $S(\pi/a)$ for different N. Hence, to simulate LEED data at room temperature using supercells of size N=30 (our computational limit), we use Monte Carlo results from higher temperatures.

In Fig. 4 we show simulated LEED spot profiles, obtained as an ensemble average (over 50 Monte Carlo configurations) of the (0,k) fractional-order intensities. The peaks for integer values of k are very sharp above and below $T_c$, since they originate from the ideal 3×1 Si substrate. For half-integer values of k, the peaks change dramatically with temperature: above $T_c$ (Fig. 4a) they are broadened and greatly attenuated, while below $T_c$ (Fig. 4b) they are sharp, with magnitudes comparable to the integer-k peaks. Thus we conclude that low-temperature LEED should provide a sufficiently sensitive test of our predictions.

We comment here briefly on an earlier theoretical proposal by Schäfer *et al.* [15], which ascribed the experimental absence of sharp LEED peaks at half-integer k to the loss of correlation between different adsorbate rows, each of which was assumed to be individually ordered. Our present results both confirm that earlier work (in that the adsorbate rows act as 1D systems with no significant correlations between them) and enlarge upon it (by revealing an additional mechanism for the loss of order within each row).

Finally, we demonstrate that at physically relevant temperatures our adsorbate model is equivalent to a long-range antiferromagnetic Ising model at fixed magnetization. Our general 1D model allows for continuous adsorbate positions $x_i$, but for temperatures of order $T_c$ the adsorbates remain very close to the $T_4$ and $H_3$ sites. By analyzing many (on the order of 10) equilibrium configurations at temperatures above $T_c$, we conclude that the loss of long-range order is due not to larger thermal fluctuations of the adsorbates about their minima. Instead, we find that two types of defects occur with increased frequency: the absence of an adsorbate from a nominally occupied $T_4$ or $H_3$ site ("vacancy"), and an adsorbate occupying a nominally unoccupied $T_4$ or $H_3$ site ("interstitial"). Hence, we consider the adsorbates to sit precisely at the local minima of the $V_{sub}(x)$ in Eq. (1), the $T_4$ and $H_3$ sites. In the Ising model, the occupied sites are represented by down spins and the unoccupied sites by up spins. Since only 1/4 of the sites are occupied, the magnetization of the model is fixed at m=1/2. The energy difference between the $T_4$ and $H_3$ sites results in a small staggered magnetic field. The Ising Hamiltonian can then be shown to be $H = \sum_{IJ} J(X_I - X_J) \sigma_I \sigma_J + h \sum_I (-1)^I \sigma_I$ where the sums are over consecutive integers labeling the $T_4$ and $H_3$ sites with positions $X_I = I(a/4)$, the Ising variables are $\sigma_I = -1$ (+1) for occupied (unoccupied) sites, and the staggered field is $h = [V_{sub}(H_3) - V_{sub}(T_4)]/2 \approx 18$ K (and hence is not significant for temperatures of order $T_c$). The long-range $J(R) = (Q^2/4\varepsilon)|R|^{-1}$ is positive, indicating antiferromagnetic interactions. Using the same statistical methods described above, we find this Ising model has a phase transition at 300 K, in good agreement with our result for the full model, $T_c$=270 K.

We note that the above Monte Carlo result is considerably lower than predicted by simple mean-field theory. There are several sources for this reduction. (1) The long-ranged antiferromagnetic interaction is frustrated. (2) Since the magnetization is constrained to be 1/2, the actual ordering occurs at a wavelength of four sites, which is twice the wavelength of the ordering in the unconstrained system. (3) Neighboring spin flips (equivalent to moving the Ba from $T_4$ to $H_3$ sites) remain, giving the model additional avenues for disordering that are not present in the simple model. (4) In general, fluctuations have dramatic effects in 1D that are not included in the mean-field result.

In summary, we have shown that the Ba-induced Si(111)3×2 reconstruction is a physical realization of a one-dimensional antiferromagnetic Ising mode with long-range interactions. Monte Carlo simulations using a more general Coulomb-gas model reveal a phase transition at finite temperature.

Acknowledgements

We are grateful to N. F. Samatova for kindly providing us with the parallel version of SATLEED. Computations were performed at the DoD Major Shared Resource Centers at ASC, ERDC, and NRL. This work was supported by the Office of Naval Research.

FIGURE LEGENDS

FIGURE 1. Fully relaxed geometry for the honeycomb chain-channel model of Ba/Si(111)3×2. Black circles are Ba, gray circles are Si. Labels **T** and **H** denote the $T_4$ and $H_3$ adsorption sites, respectively.

FIGURE 2. Variation of the DFT total energy for relative displacements of a Ba adsorbate between 0.00 ($T_4$ site) and 1.00 ($H_3$ site). Circles: all Ba displaced rigidly. Squares: every second Ba held fixed at the $T_4$ site. The functional forms of the curves are described in the text.

FIGURE 3. (**a**) Specific heat computed from Monte Carlo simulations for the 2D model (heavy curve) and its 1D analog (light curve), with N=25 for both. (**b**) Structure factor for the 2D model with N=25 (heavy curve), and its 1D analog with N=25 (light curve), N=50 (dotted), N=100 (dashed), and N=200 (dot-dashed). (**c**) Finite-size scaling of the critical temperature, predicted from the fourth-order cumulants, as a function of average system size in 2D (squares and heavy line) and 1D (circles and light line).

FIGURE 4. Simulated LEED profiles at (**a**) 1000 K; (**b**) 200 K.

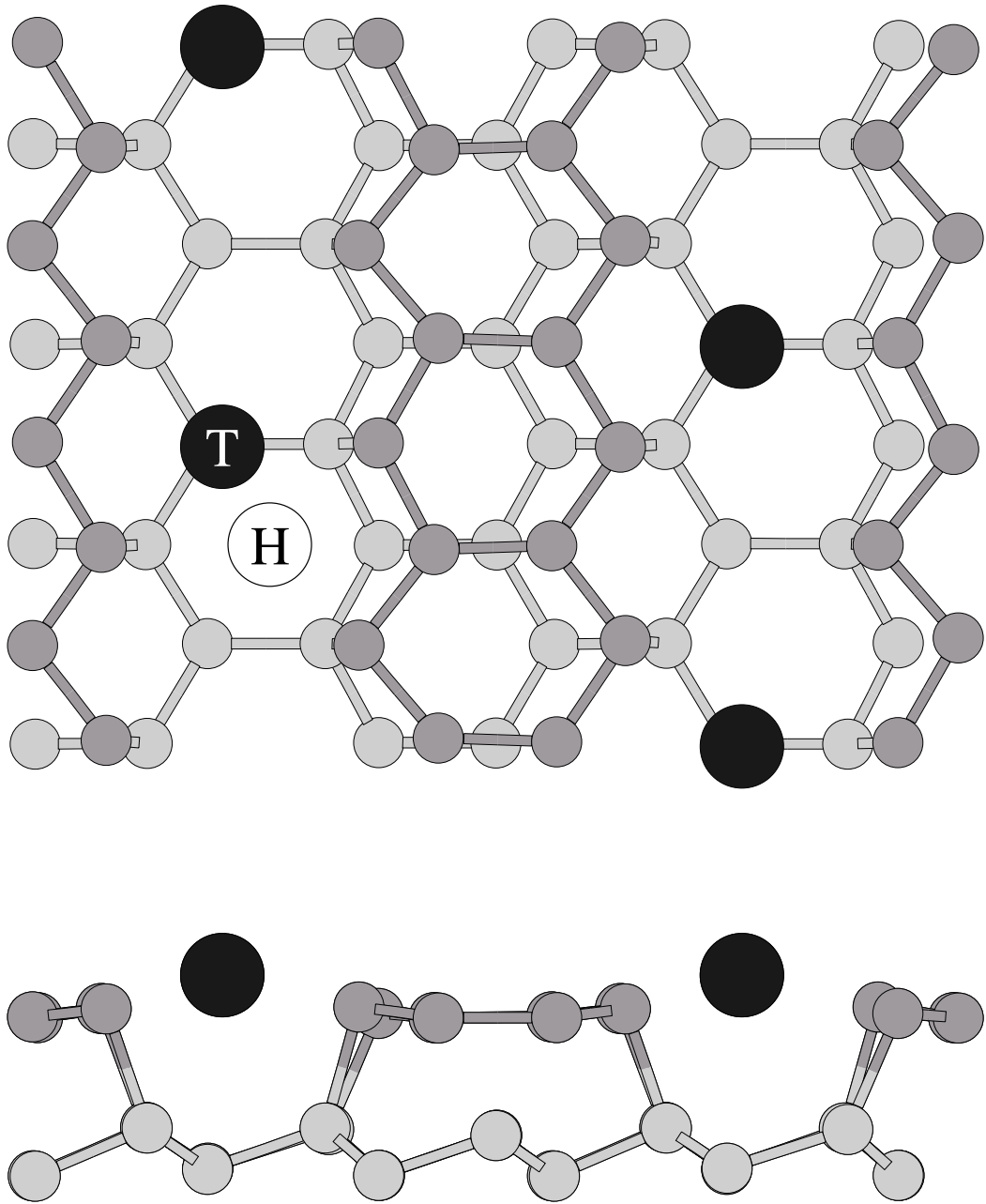

FIGURE 1

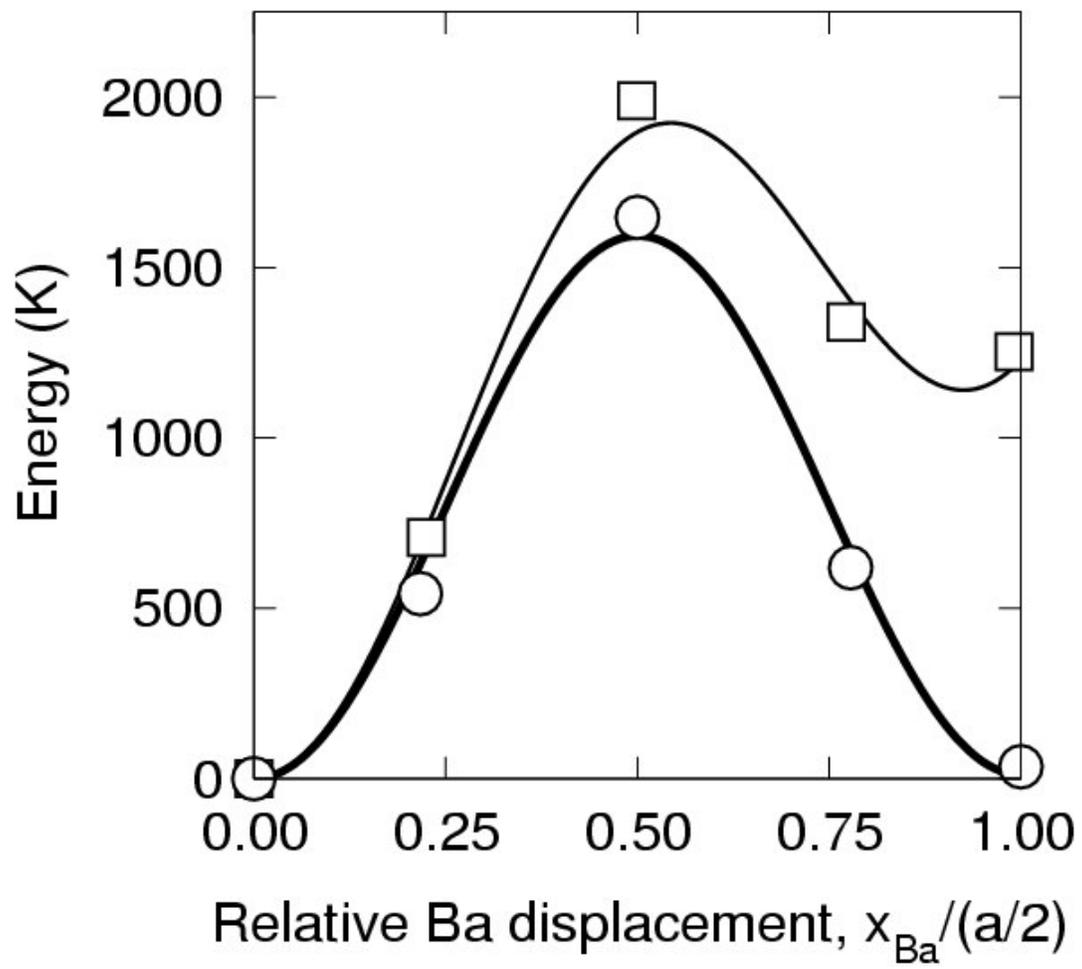

FIGURE 2

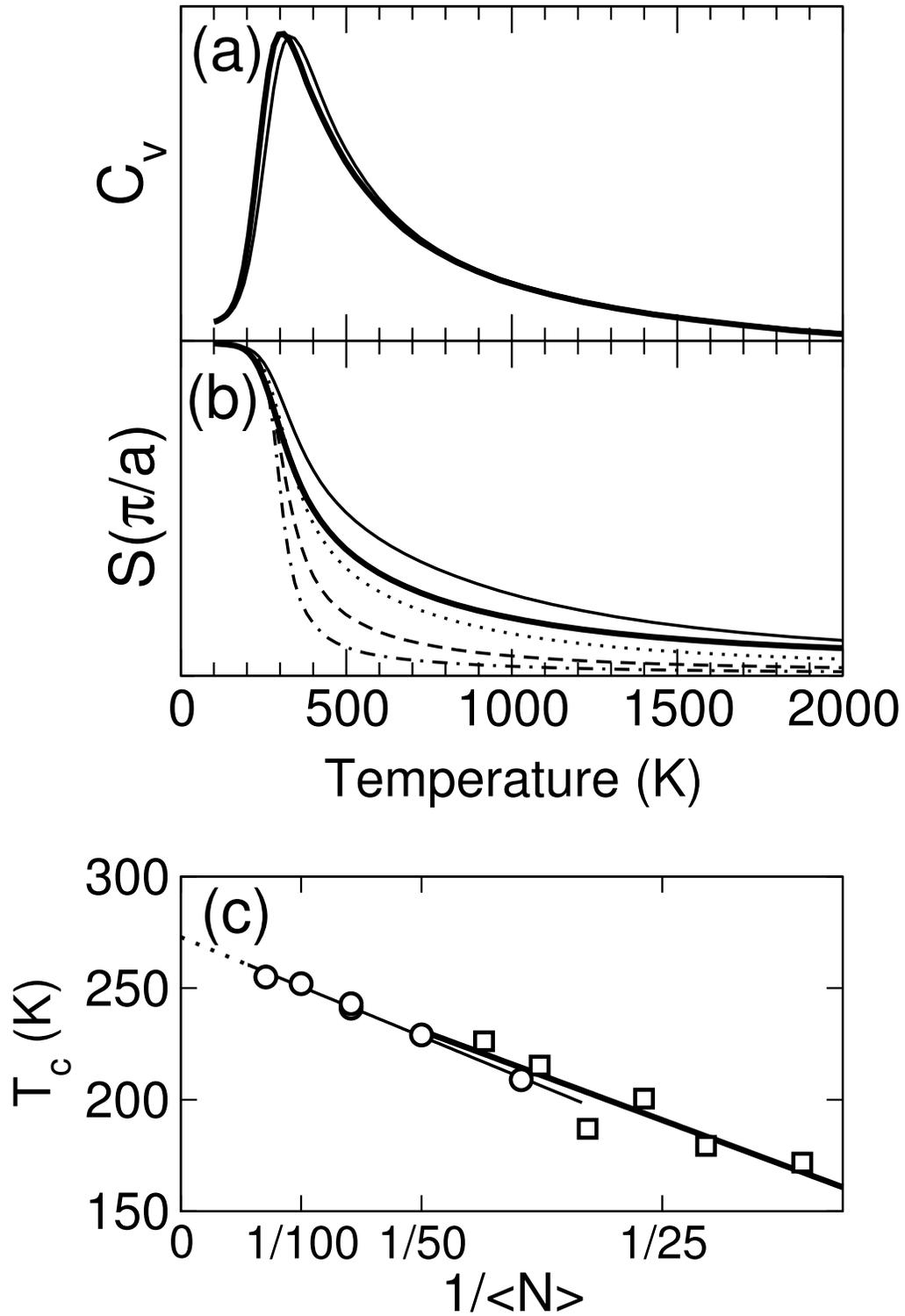

FIGURE 3

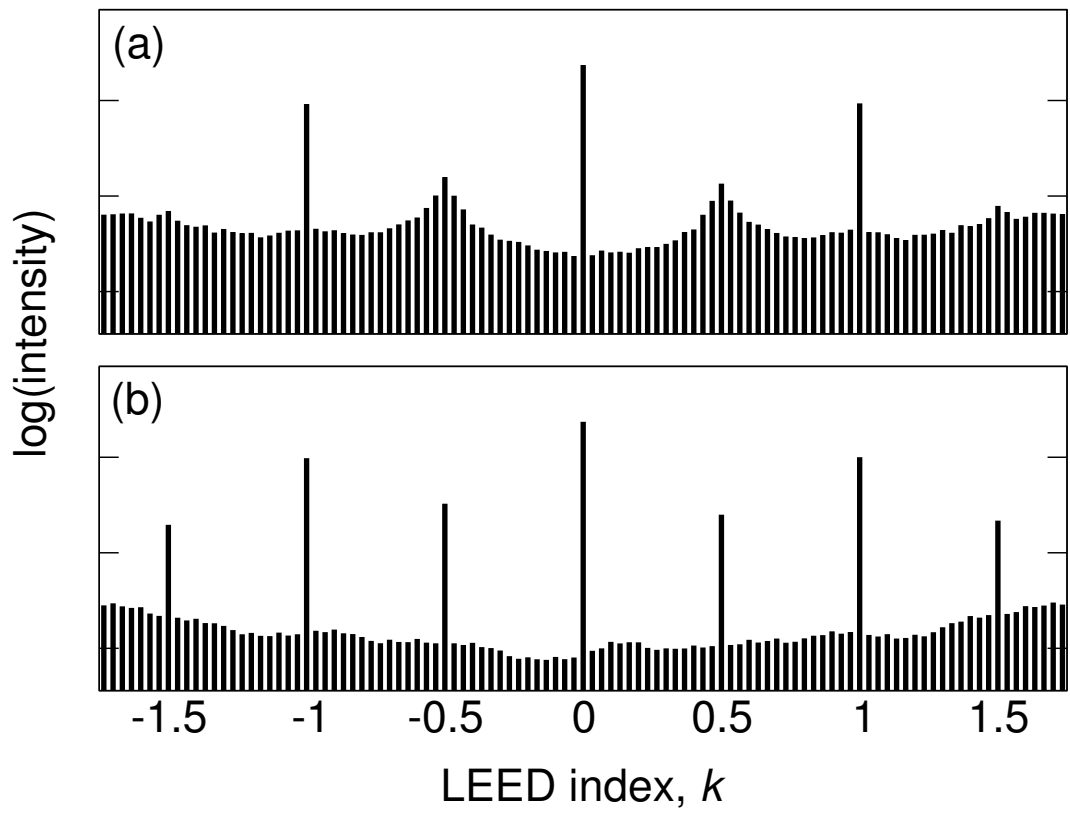